# Modeling the Realization and Execution of Functions and Functional Requirements

Sabah Al-Fedaghi
Computer Engineering Department
Kuwait University
Kuwait
sabah.alfedaghi@ku.edu.kw

*Abstract*—Requirements engineering plays a critical role in developing software systems. One of the most difficult tasks in this process is identifying functional requirements. A critical problem in many projects is missing requirements until late in the development cycle. In this paper, our core interest is function modeling, which refers to building models of systems based on their functionalities and on the functionalities of their subcomponents. We present a framework as the basis for specifying functional requirements via a modeling language that produces a high-level diagrammatic representation. The aim is to deliver an overall system description, facilitate communication and understanding, construct a holistic view of the system above the domains of different expertise, and lay the foundation for the design phase. We analyze the notion of function and its elementary types and apply examples of natural language description and scenarios. The results reveal a new method that lays a foundation for works on functionality and viable methodology for capturing its requirements.

*Keywords-function; functional requirement; conceptual model; reqirement modeling, elementary functions*

## I. INTRODUCTION

One of the most difficult tasks when developing software is identifying functional requirements [1], which prescribe a system's desired functionality [2]. Shortcomings in the treatment of requirements are the main cause of software projects' failure [3-5]. Accordingly, having a clear understanding of functional requirements and describing them in an accurate and unambiguous manner is a necessity ([6], as cited in [7]).

A requirement is "a statement of need" [5] or, as defined by [8]

- A capability the user needs to solve a problem and achieve an objective
- A capability that a system must possess to satisfy a specification.

Functionality refers to features that satisfy a user's needs. According to [9], functional knowledge (in contrast to physical and behavioral knowledge) is present in all domains of life. A function is something a system does; according to [5], the common term "functional requirement" in practice means the same as function.

In this paper, our core interest is "function modeling", which refers to developing models of systems (e.g., devices, products, objects, and processes) based on their functionalities and on the functionalities of their subcomponents [10]. Constructing such a high-level representation model provides an overall system description to facilitate communication and understanding between engineers of various disciplines. Function modeling mainly concerns how to represent knowledge about function. It also provides "a holistic view of the system above the domains of different expertise and serves as a means of linking the upper and lower levels of system design and description" [10]. We focus on representation issues regarding knowledge accumulated during the requirements acquisition phase of system development, a topic referred to as requirements modeling [2].

In this context, modeling denotes systematically identifying the domain's relevant aspects [2] to directly represent these aspects in the requirements. We focus on *conceptual* modeling: the activity of describing features of the physical and social worlds around us to facilitate understanding and communication [11-12].

Specifically, we present a framework (underlying concepts, operations, and structure) that forms the basis for modeling requirements. The involved modeling language *Thing Machine* acronymized as "TM". For this goal, we look at the fundamental notions of function and requirements, and then we express some conceptualizations of function in TMs. Also, we produce TM-based models for the given functional requirements of systems in the software literature.

### A. Function: Significance and Problem

Function is "a central concept in common-sense and engineering descriptions of artefacts" [13]. According to Socrates, each thing has a function. Aristotle claimed that we must identify the function of a human being to discover the human good. A human's function is rational activity [14]. In modeling and designing, knowledge of functions is important for visualization, explanation, evaluation, diagnosis, and repair of designs and processes [15-17]. The idea of function is central to engineering, where design artifacts are intended to have certain functions: "When designers are designing, they look for components that can achieve certain functions. Predicting how systems would behave under various conditions of use or abuse also often requires knowing what the functions of the devices are. Thus, the notion of function is certainly of fundamental importance in engineering" [18]. Nevertheless, "there is a lack of a domain-independent framework for representing and modeling functions" [9]. According to [19], despite many years of research and wide use in various domains and research areas, several issues exist with the notion of function. Researchers attribute various meanings to function and sometimes use it interchangeably with behavior [20], purpose, and operation [21].

Several issues hinder progress with understanding and using functions, including the lack of a clear and overarching definition [22]. Function lacks a precise meaning, and its various ambiguous descriptions create confusion in communication and archiving and obstruct teaching and formalization [19,23].

### B. A New Approach to Functions and Functional Requirents?

In this paper, we will discuss such questions and issues via a new approach to conceptual modeling [24-35]. The next section summarizes related works about functionality. To develop a reasonably self-contained paper, Section 3 reviews basic notions of TM, as mentioned previously. The subsequent sections apply TM to many functions and functional requirements.

## II. THE NOTION OF FUNCTION AND RELATED WORKS

In many domains, entities or processes are considered in terms of their functions [9]. This section briefly reviews functionality and related works in computing.

### A. What is Function?

"Function's" meaning is widely used but rarely defined [36]. This highly ambiguous term [37] has been discussed at length in both philosophical and biological circles [38]. The Oxford English Dictionary describes "function" as "an activity that is natural to or the purpose of a person or thing".

"Function" appears to be the name for a set of related ideas rather than for a single concept [18]. It may have many different meanings:
- Function should mean why an entity does what it does.
- what an entity does describes the contribution of an entity to a complex system [39]

Reference [5] further defines a function as "something that a system or subsystem does, presumably because a requirement made it necessary". They also state that the commonly used term "functional requirement" means "in practice. . . the same as function" [19]. Reference [40] defines function as the transformation from input to output; input and output can be types of material, energy, or signals. Reference [41] viewed an activity as any function or process that transforms inputs into outputs. Function has also been defined as a change designers implement between two scenarios: before and after the design's introduction [22].

Reference [42] expresses function as "to do something". The function is represented using a verb and a noun. Reference [42] distinguishes between primary and secondary functions. For example, a domestic pump's primary function is to pump water, and its secondary function is to operate with little noise. Reference [43] considers functions as fulfilling the expectations of the resulting artifact's purposes. Reference [44] defines function as an artifact's intended purpose. The function describes the artifact's goals at a level of abstraction that is of interest at the artifact level. Reference [3] distinguished between function (relating to purpose) and behavior (describing actions). A thing is generally understood to have various functions according to its position in a system—e.g., a stone may be function as a weapon, a meat-cutter, or a brick in a wall.

### B. Functionality in Computing

Computer science includes vast research in artificial intelligence concerned with functional representation in the context of design—e.g., what an artifact is supposed to do, which is its function. Reference [9] reviewed many works in this area and showed a strong correlation between function and behavior. Functional representations in this context have different interpretations of a function in terms of input–output, effects, and intended roles [9].

In software engineering, functional decompositions and functional dependencies are familiar approaches in modeling systems. Functionality is understood as a software system that meets requirements; if it does not meet requirements, then we say it does not perform its function [9]. In the development process, functions mostly relate to three phases: the requirements/design phase, the implementation/development phase, and the verification/testing phase [45]. ISO2476510 [46] relates function to the requirements phase. Additionally, the term "functional requirement" is viewed as a requirement that specifies a function that a system or system component must be able to perform. "Functional design" is a term used to denote the process of defining working relationships among a system's components and the result of that specific process. Functional testing [46] is considered black-box testing that ignores the system's internal mechanisms and focuses on outputs generated in response to selected inputs and execution conditions.

In object-oriented modeling, functionally is forced into object form [47], where each object contains functionality that pertains to it. In object-oriented programming, the functionality is encapsulated into the class methods [48]. As an alternative approach to object-orientation, the object-process methodology (OPM) [49] defines a function as an object's attribute (e.g., your heart's function is to pump blood). Such a function describes "the intent for which it [object] was built, the purpose for which it exists, the goal it serves, or the set of phenomena or behaviors it exhibits" [49]. For Dori [49], the function (the what and the why) and dynamics (the how) are distinct concepts [9].

## III. THINGING MACHINE

This section will briefly review thinging machine modeling to provide a base for this paper's aim of taking TM as a foundation to study functionality. A more elaborate discussion of TM's philosophical foundation can be found in [26-29].

### A. TM Model

One question first raised in modeling is ontology: What kind of things are models? The TM ontology is based on a single category of entities called thimacs (*thi*ngs/*ma*chines). The thimac is simultaneously an "object" (called a *thing*) and a "process" (called a *machine*)—thus, the name thimac. The thimac notion is not new. In physics, subatomic entities must be regarded as particles and as waves to fully describe and explain observed phenomena [50].

According to Sfard [51], abstract notions can be conceived in two fundamentally different ways: structurally, as objects/things (static constructs); and operationally, as processes. Thus, distinguishing between form and content and between process and object is popular, but, "like waves and particles, they have to be united in order to appreciate light" [52]. Processes and objects must unite in order to understand modeling.

TM adopts this notion of duality in conceptual modeling, generalizing it beyond mathematics and its utilization in software engineering modeling. "Structural conception" means seeing a notion as an entity with a recognizable static structure. The operational way of thinking emphasizes the dynamic process of performing actions. A model describes a given domain independent of technological choices that could impact the implementation of a system based on itself.

The term "thing" relies more on Heidegger's [53] notion of "things" than it does on the notion of objects. According to Heidegger [53], a thing is self-sustained, self-supporting, or independent—something that stands on its own. A thing "things". That is, it gathers, unites, or ties together its constituents in the same way that a bridge unifies environmental aspects (e.g., a stream, its banks, and the surrounding landscape).

The term "machine" refers to a special abstract machine called a "thinging machine" (see Fig. 1). The TM is built under the postulation that only five generic processes of things are performed by a TM: creating, processing (changing), releasing, transferring, and receiving. Informal justification for limiting modeling to these five generic (primitive/elementary) operations is given in [31].

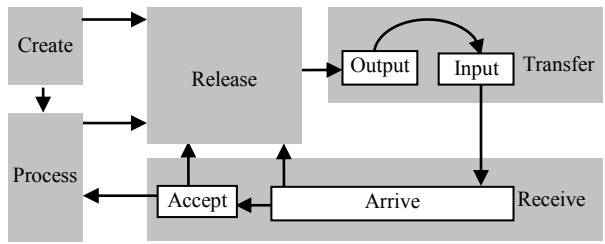

**Figure 1. A thinging machine.**

A thimac has dual being as a thing and as a machine. A thing is created, processed, released, transferred, and/or received. A machine creates, processes, releases, transfers, and/or receives things. We will alternate between the terms "thimac", "thing", and "machine" according to the context.

Ontology requires classifications, such as a functional classification of human body functions: mental, sensory, speech, respiratory, digestive, and so on. [54]. Even with impressive progress in developing ontologies of things (i.e., entities, objects), the ontology of processes (TM machines) is still a problem [54].

The five TM operations (also called stages) form the foundation for thimacs. Among the five stages, flow (a solid arrow in Fig. 1) signifies conceptual movement from one machine to another or among a machine's stages. The TM stages can be described as follows.

- *Arrival*: A thing reaches a new machine.
- *Acceptance*: A thing is permitted to enter the machine. If arriving things are always accepted, then arrival and acceptance can combine into the "receive" stage. For simplicity, this paper's examples assume a receive stage.
- *Processing* (change): A thing undergoes transformation that changes it without creating a new thing.
- *Release*: A thing is marked as ready to be transferred outside of the machine.
- *Transference*: A thing is transported somewhere outside of the machine.
- *Creation*: A new thing is born (created) in a machine. A machine creates in the sense that it "finds/originates" a thing; it brings a thing into the system and then becomes aware of it. Creation can designate "bringing into existence" in the system because what exists is what is found.

In addition, the TM model includes memory and triggering (represented as dashed arrows) relations among the processes' stages (machines).

In TM, the world is abstracted as thimacs with five generic stages. A thimac may contain a single stage, but no create, process, release, transfer, and receive stage can be separate from a thimac. This saying makes a good analogy: "The hand separated from the body is not a true hand" (Aristotle, *Politics*).

The grand thimac is not a single monolithic, unmanageable whole, but incorporates decomposability by its skeletal structure of multiple interior thimacs. This decomposability is based on joints (flows and triggering among thimacs) that form the structure (anatomy) of a system (the overarching thimac). The joints are gaps that separate sub-functionality. Such a system's dynamism is described by embedding time into internal thimacs. Accordingly, the conceptual model is a single diagram, but the implementation (e.g., software) lends itself to carving thimacs at the joints via an adequate conception of events. According to Plato, the world (in TM, the grand thimac) comes to us predivided with joints to carve [55].

### B. General Idea of TM-Based Functionality

We mentioned previously that one of the definitions of a function is something that a system does. This paper's main idea is that "doing" in a TM is expressed by the creating, processing, releasing, transferring, and receiving embedded in a thimac. Thus, the TM model is the realization (in terms of the TM's static description) and execution (in terms of a TM's dynamic description) of the system's function. This includes the system as a thing and as a machine (process).

The system (as the grand thimac) makes its function accessible (e.g., to designers, users) by unfolding its operations (create, process, release, transfer, and receive), flows, and events that must be performed to fulfill those functions. Reference [56]'s description of technology means the grand thimac (system) is forced to reveal itself (and hence, its functionality) by the TM's modeling methodology.

More simply, functionality is a web of interdependent thimacs that can engage in complex behaviors to bring about the grand thimac's behavior. Each piece of this web of thimacs has its own sub-function that contributes to the requirements of other thimacs and to the grand thimac's function. The environment of grand thimacs may supply flows and triggers.

## IV. APPLYING TM TO EXAMPLES

This section applies these ideas to the notion of functionality.

### A. An Automobile's Function

An automobile's function is transporting people and goods [57]. Fig. 2 shows a thimac that describes this. Fig. 2 models only the structure of the realization of function that grounds the function's capacities. To model its execution in terms of behavior, the notion of "event" is incorporated into the TM. An event encompasses at least three subthimacs: time, region (of event), and the event itself. For example, Fig. 3 shows this event: *The automobile moves from place 1 to place 2*. For simplicity, we will represent each event only by its region.

Accordingly, we can decompose the TM description of the automobile as follows (see Fig. 4).

- Event 1 ($E_1$): Things move into the automobile at place 1.
- Event 2 ($E_2$): The automobile moves from place 1 to place 2.
- Event 3 ($E_3$): Things move out of the automobile in place 2.

Fig. 5 shows the chronology of these events. This idea will be explored further by more complex examples in this section.

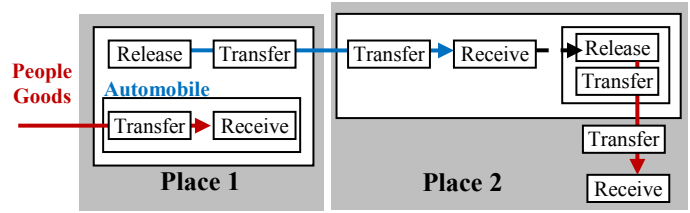

**Figure 2. The function of an automobile as a thimac.**

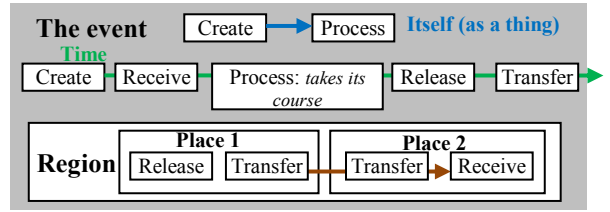

**Figure 3. The event "the automobile moves from place 1 to place 2."**

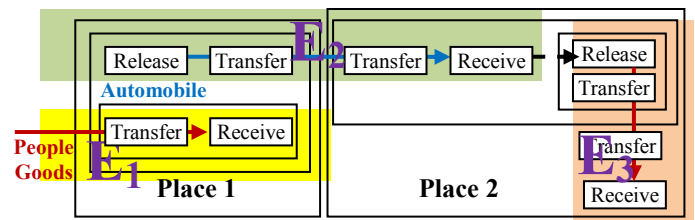

**Figure 4. The events in function of an automobile as a thimac.**

$$E_1 \rightarrow E_2 \rightarrow E_3$$

**Figure 5. Chronology of events in function of an automobile as a thimac.**

### B. Function Decomposition

Engineers construct models that represent the structure of artifacts' structures. According to [23], "They managed to model the structure but the incorporation of functional decomposition is less successful." Functional decomposition means dividing a larger task (thimac) into smaller tasks (subthimacs). Decomposability refers to logically grouping a system's components into subsystems.

For example, according to [19], a coffee mill's function is to convert coffee beans, electrical energy and electrical signals into coffee powder, heat energy, and electrical signals. As seen in Fig. 6, coffee beans flow to the coffee mill, where they are processed to trigger the creation of coffee powder. Fig. 6 can be decomposed according to its dynamic behavior, which is specified by events. Accordingly, we can decompose the coffee mill's TM description according to its events as follows (see Fig. 7).

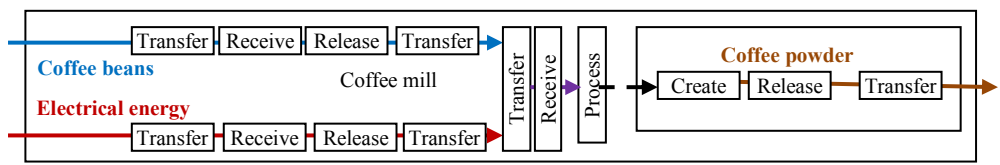

**Figure 6. Function as a transformation from input to output [19].**

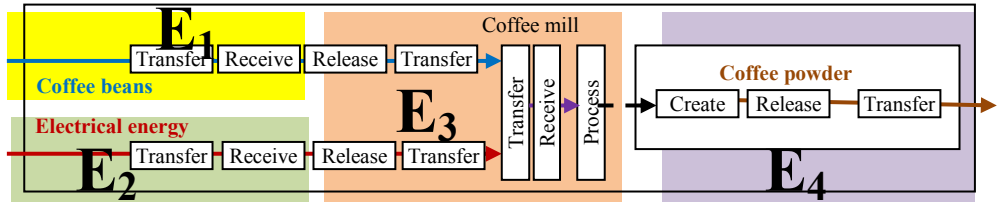

Figure 7. The events in the mill TM representation.

- Event 1 ($E_1$): Getting the coffee beans
- Event 2 ($E_2$): Providing electricity
- Event 3 ($E_3$): Processing the coffee beans using electricity
- Event 4 ($E_4$): Creating coffee powder

Fig. 8 shows these events' chronology. These four events execute the functionality of the coffee mill's four parts. The general functionality can be extracted from these sub-functions.

Reference [19] stated a coffee mill's function as converting coffee beans, electrical energy and electrical signals into coffee powder, which can be stated as follows.

- Processing coffee beans when the mill gets coffee beans and is provided with electricity
- Producing (creating) coffee powder.

The first part corresponds to "Convert coffee beans, electrical energy and electrical signals," and the second part corresponds to "into coffee powder". A better rephrasing of the function of a coffee mill is "processing coffee beans using electricity to produce coffee powder."

Accordingly, we can describe the coffee mill's functionality as illustrated in Fig. 9. This explanation aligns with [58]'s account of a sub-function as a component part's contribution to the whole function of a system that contains that component. The function of a part references the reasons that part was incorporated into the system.

We can conclude that TM modeling provides a systematic method to specify the realization (static description; Fig. 6), the execution (dynamic description; Figs. 7 and 8) and functionality of the coffee mill.

### C. Primary and Secondary Functions

Reference [42] distinguishes between primary and secondary functions. For example, a domestic pump's primary function is to pump water, and its secondary function is to operate with little noise.

Fig. 10 shows this pump's TM model, and Fig. 11 shows its events: receiving water, processing water, and creating pressured water. The event diagram shows that this functionality can be accomplished without event $E_4$, so event $E_4$ is nonfunctional. Thus, this notion of secondary function is not applicable in [42]'s example.

We conclude that the TM model clarifies a system's functionality. In the given example, the noise is a byproduct of the pump's processing part. Its low level of noise is not a functional requirement. If the pump functionality is defined as receiving water, processing water with little noise, and creating pressured water, then little noise is part of the functional requirements as shown in Fig. 12.

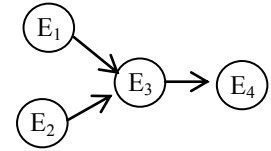

Fig. 8. Chronology of events in the coffee mill example.

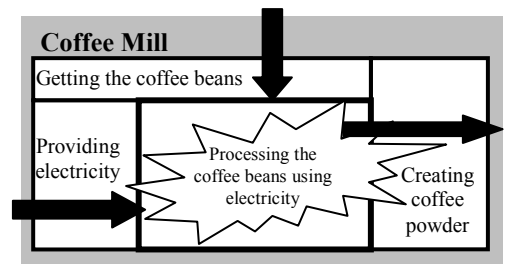

Figure 9. Decomposition of the mill's function.

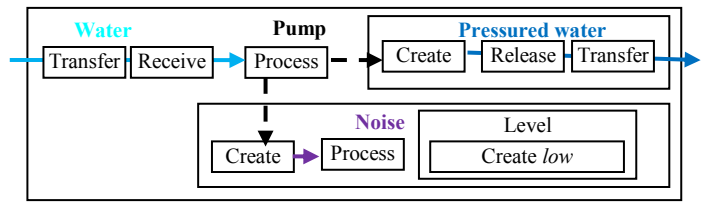

Figure 10. TM model of the pump example.

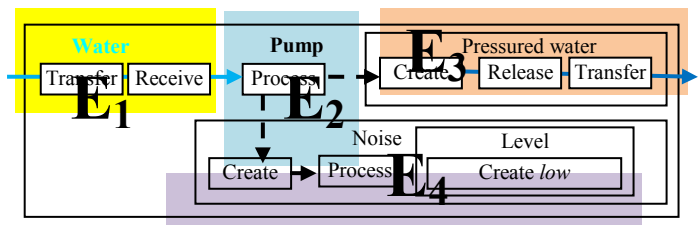

Figure 11. Events of the pump example's TM representation.

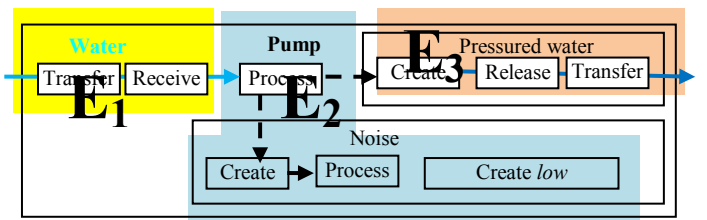

Figure 12. The pump example after eliminating the so-called secondary function.

*D. Independent Functions*

Reference [43] considers functions as fulfilling the expectations of the resulting artifact's purposes. When designing windows, for example, some functions are providing daylight and controlling ventilation. In his design prototype schema, [43] represents functions using combinations of verbs, nouns, and adjectives (e.g., provide daylight, control ventilation).

Alternatively, Figs. 13 and 14 show the window example's TM model. We can see that the TM may represent several independent functions.

*E. Function of a Machine and Parts of a Machine*

According to [18], objects and processes both have functions. Boiling is a process that can create steam. Distillation is a process that can separate components.

These two processes are represented in Figs. 15 and 16. The figures show boiling and distilling (presumably water) as thimacs (machines). Fig. 15 includes the heat process needed by the system. Boiling is a thimac that receives, processes, and creates steam and outputs it. Similar wording can apply to the distillation example.

Thus, the function is not the function of a mere "process", but it is the function of the boiling and distillation machines. Reference [18] conceptualized a process as an operation of change, not of creation in the TM sense (as shown in Fig. 17).

## V. FUNCTIONAL REQUIREMENTS

This section continues to apply TM to functionally, but in the context of larger issues: sharing functionality, functional requirements, and relationships among sets of functional requirements.

*A. Sharing Functionality*

Requirements are generally specified using textual specifications and use cases. In reality, the use cases do not often meet the standards of a "good" use-case model (see [7] and its sources). Reference [7] raised the issue of reuse of shared functionality (e.g., via UML relationships for use cases) and indicated that this problem is "seldom employed in practice." Rago et al. [7] argued that the duplication of functionality in use case specifications is a major problem in software development. Duplicating functionality is the action of repeating the descriptions.

Researchers usually focus on textual requirements, such as use cases with duplicate pieces of functionality scattered across the textual specifications. According to [7], "Duplicate functionality can sometimes improve readability for end users, but hinders development-related tasks such as effort estimation, feature prioritization and maintenance, among others. Unfortunately, inspecting textual requirements by hand in order to deal with redundant functionality can be an arduous, time-consuming and error-prone activity for analysts."

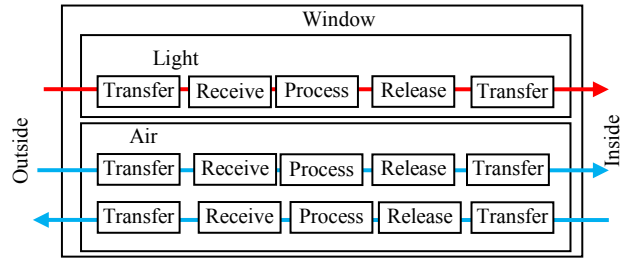

**Figure 13. Some functions of a window are provision of daylight and control of ventilation [43].**

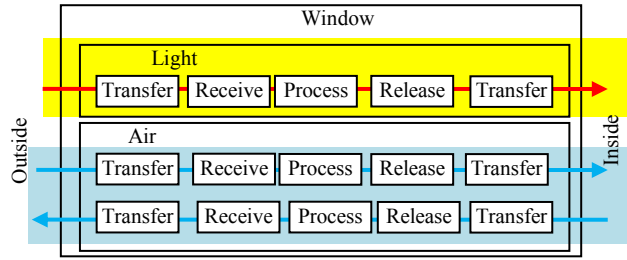

**Figure 14. Events: provision of daylight and control of ventilation.**

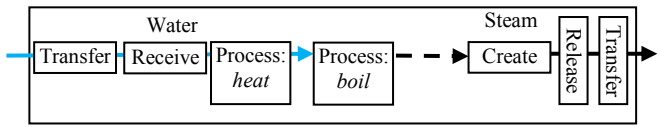

**Figure 15. Boiling is a process that can be used to create steam [18].**

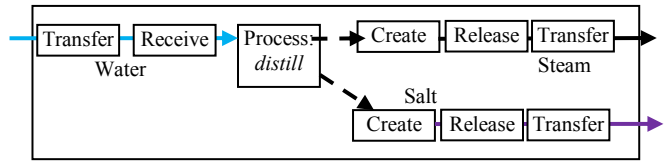

**Figure 16. Distillation is a process that can be used to separate components [18].**

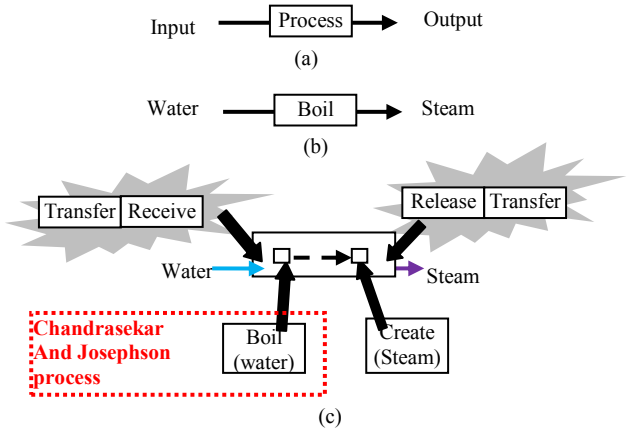

**Figure 17. (a) Classical Input-Process-Output model, (b) Chandrasekar and Josephson [18] model, and (c) TM model.**

Reference [7] introduced an approach that helps analysts automatically spot signs of duplication in use cases. Their method combines several text-processing techniques, such as a use-case-aware classifier and a customized algorithm for sequence alignment. The use cases are converted into an abstract representation comprising sequences of semantic actions, and these sequences are then compared pairwise to identify action matches, which become possible duplications.

Reference [7] gave an example of an automatic teller machine with five use cases, namely Transfer Money, Withdraw Money, Deposit Money, Pay Service, and Add Service. Reference [7] studied the resemblance between the Pay Service and the alternative flow of Add Service, which constitutes a duplicate behavior.

The text of the involved use case is as follows.
1. The system enumerates the kind of service available.
2. The client chooses a kind of service.
3. The system lists companies providing that kind of service.
4. The client chooses a company.
5. The system requests the client's service identifier code.
6. The client enters the code and confirms the operation.
7. The system records the new service.
8. The use case ends.

The same scenario occurs as an alternative flow in the Pay Service use case (see Fig. 18).

Fig. 19 shows the TM representation of the above scenario. First, the system downloads the list of the kinds of services (numbers 1 and 2 in the figure). The client processes the list to trigger (3) the creation (4) of his/her selection, which flows to the system (5), and so on.

The TM diagram can clearly be converted to a graph. Fig. 19 can be simplified by removing Release, Transfer, and Receive under the assumption that the arrow's direction indicates the direction of flow, as show in Fig. 20. Further simplification makes it easy to produce the graph (a network of nodes with edges connecting some nodes) shown in Fig. 21.

Accordingly, we can convert Pay Service and the alternative flow of Add Service to graphs and conclude that [7]'s basic problem of sharing functionality is the known graph isomorphism problem. This problem arises when two graphs are really the same graph because of a one-to-one correspondence (an isomorphism) between nodes that preserves how the nodes are connected. This graph isomorphism problem is neither known to be in P nor known to be NP-complete (see the latest results about this problem in [59]).

Thus, TM modeling can be a tool in studying shared functionality (e.g., heuristic algorithms) because the problem may depend upon the input graphs.

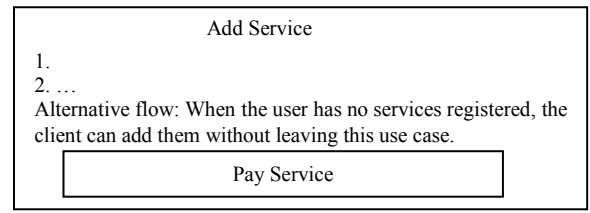

**Figure 18. The Add Service use case that includes shared functionality.**

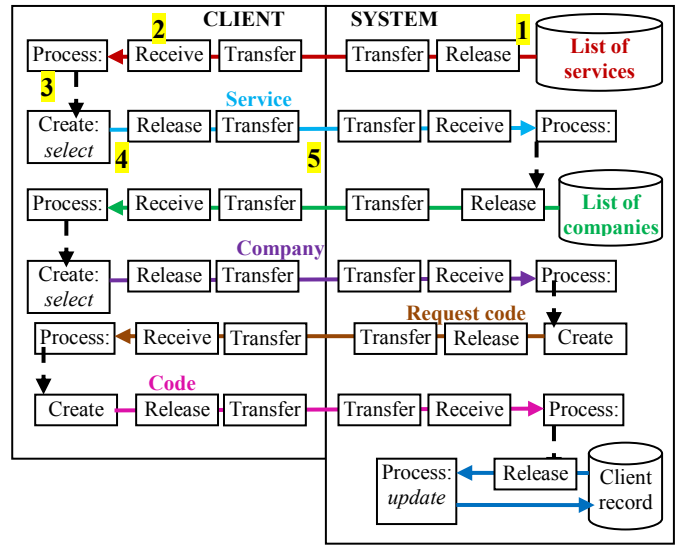

**Figure 19. TM static representation of the Pay Service.**

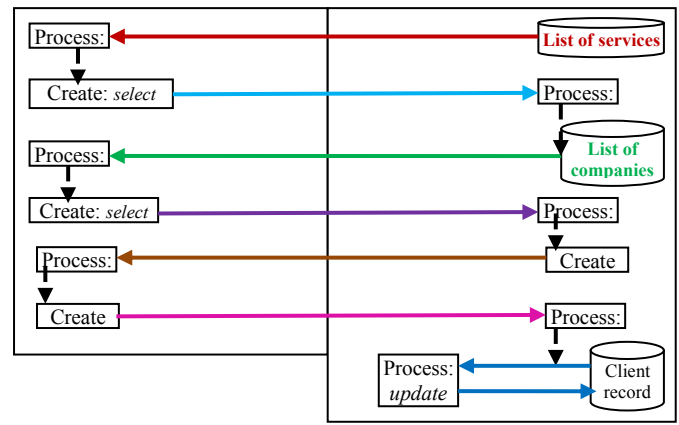

**Figure 20. Simplification of the TM static representation of the Pay Service.**

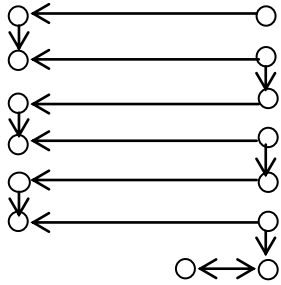

**Figure 21. The graph of Fig. 19.**

## B. Functional Requirements Quality

Many problems during the software development process relate to deficient requirement specifications. Creating specifications is a complex process because of various viewpoints and needs, which may be contradictory [60]. Scenario-based representations are frequently used to specify requirements. scenarios are usually written in natural language, so requirements engineers must focus on issues like unambiguity, completeness, consistency, and correctness. According to [60], numerous techniques have been developed to deal with quality problems in requirements specification; however, most of these techniques still do not meet desired quality levels.

Reference [60] proposed using Petri nets [61] to detect wrong information, missing information, and erroneous situations possibly hidden within scenarios. Scenarios are translated into Petri nets. Then, scenarios and their Petri nets are analyzed to identify defects that are traced back to the scenarios, allowing their revision.

Fig. 22 shows an example of using a Petri net to represent the producer and consumer problem. The problem exemplifies communicating processes using a "buffer"; the producer sends a message to the consumer. Fig. 23 shows the TM model of the producer and consumer problem. The producer and the consumer synchronize their activities as illustrated in Figs. 24 (events) and 25 (chronology of events).

Thus, a TM provides an alternative to Petri nets in developing precise requirements descriptions. Admittedly, in contrast to TMs, Petri nets already have a mathematical foundation. Nevertheless, TMs seem to provide easier presentations.

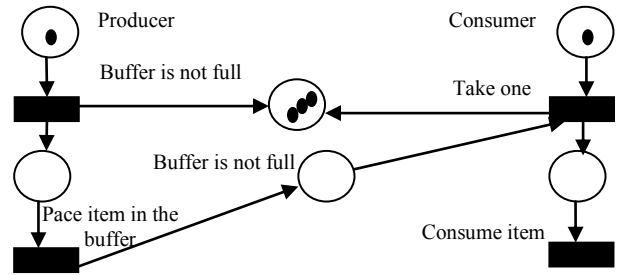

**Figure 22. Petri nets (Adapted from [60]).**

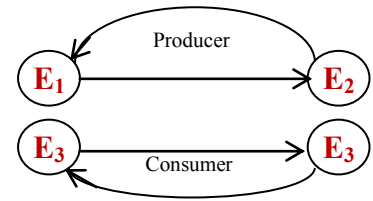

**Figure 25. The behavior of the producer and consumer problem.**

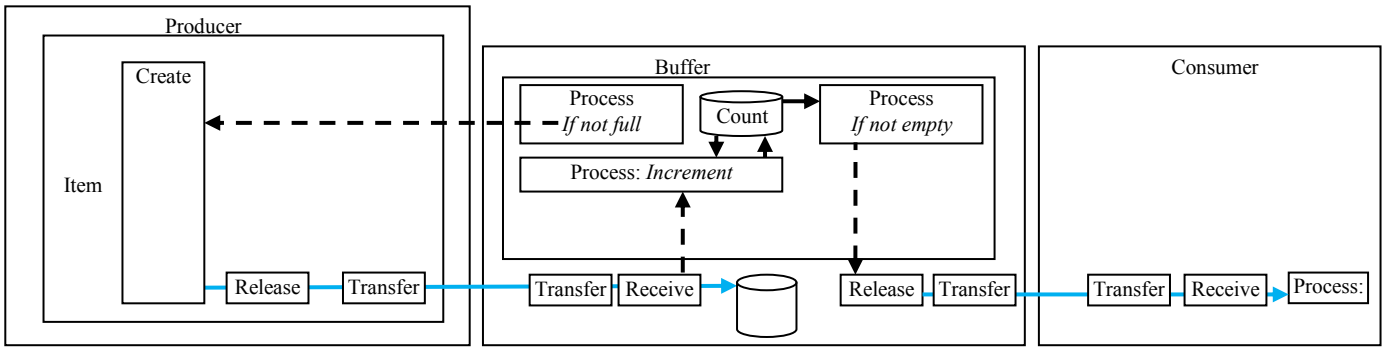

**Figure 23. The TM static representation of the producer and consumer problem.**

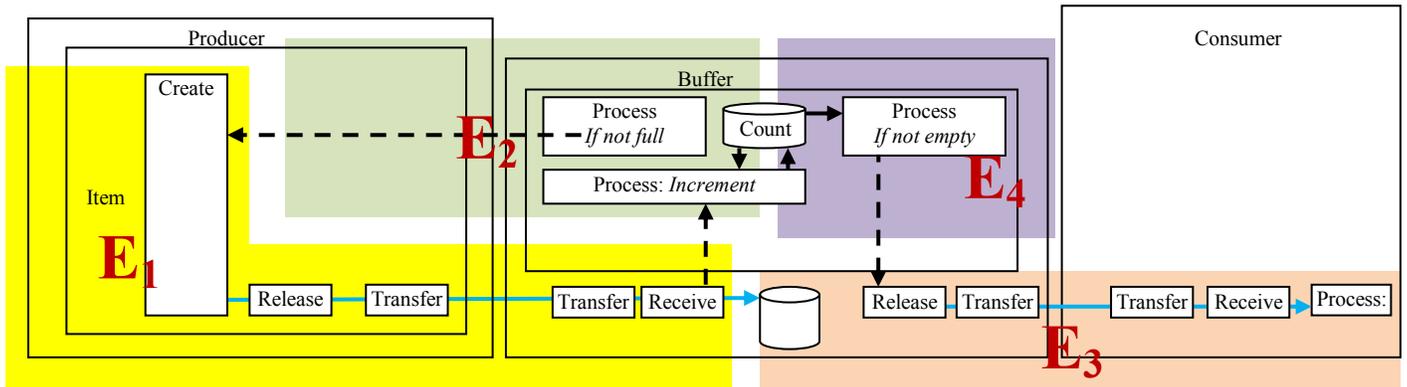

**Figure 24. The events in the TM representation of the producer and consumer problem.**

## C. Relationships Among Sets of Functional Requirements

Reference [60] re-described a Submit Order scenario of [62] as part of an Online Broker System. The Submit Order scenario is related to other scenarios by sequential and explicit non-sequential relationships. In [62]'s original version, the relationships among related scenarios are not obvious.

The problem here is that some scenarios may be specified separately, so there is a need to "connect" them in sequence. If we have the two Submit Order and Process Bids scenarios, how do we unite them?

This problem is solved by identifying where the linkage can be established. Reference [60] specifies the Submit Order scenario as follows.

1. The customer loads the login page.
2. The broker system asks for the customer's login information.
3. The customer enters her login information.
4. The broker system checks the provided login information.
5. The broker system displays an order page.
6. The customer creates a new Order.
7. DO the customer adds an item to the Order WHILE the Customer adds more items to the order.
8. The customer submits the Order.
9. The broker system broadcasts the Order to suppliers
10. # LOCAL SUPPLIER BID FOR ORDER
11. INTERNATIONAL SUPPLIER BID FOR ORDER #
12. PROCESS BIDS

From the TM perspective, the problem is caused by fragmenting scenarios into high-level functions.

Submit Order and Process Bids are high-level functions constructed from lower functions, up to the level of the TM's five generic operations. Thus, in [60]'s problem, a TM does not occur because the lines among different parts (scenarios) are established at the flow's basic level. This means there is a release/transfer/transfer/receive flow of triggering that would connect one part of the functional specification to the other part.

Fig. 26 shows the corresponding TM model. In the figure, the system downloads the login page (1) to the customer (2). The customer processes the page to create the login information (3), which flows to the system (4), where it is processed to download the order page (5). The customer receives the order page and processes it (6) to create an order (7) that flows to the system (8). The order is inserted (9) into the list of orders (10 and 11). The user signals (12) as finished ordering, so the system triggers (13) sending the list of orders to the suppliers (14). The supplier receives and processes the orders (15) to create bids (16) that flow to the system (17). The system sends the bids to the customer (18), where they are processed (19), and the customer accepts a bid.

Figs. 27 and 28 shows the events and behavior, respectively, of the Submit Order system.

## VI. CONCLUSION

In this paper, we presented a framework that formed the basis for functional modeling language as a high-level diagrammatic representation. The aim is to deliver an overall system description to facilitate communication and understanding, constructing a holistic view of the system above the domains of different expertise and to lay a foundation for the design phase.

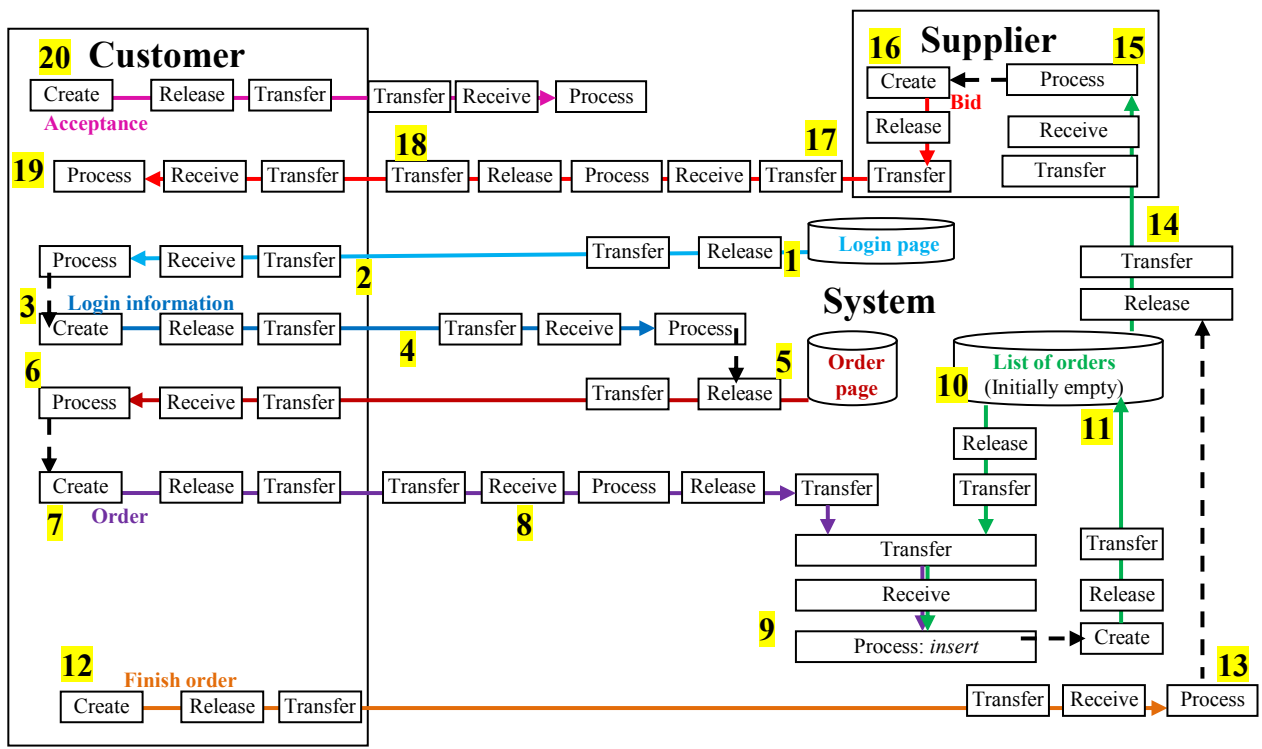

Figure 26. The TM model of Submit Order scenario

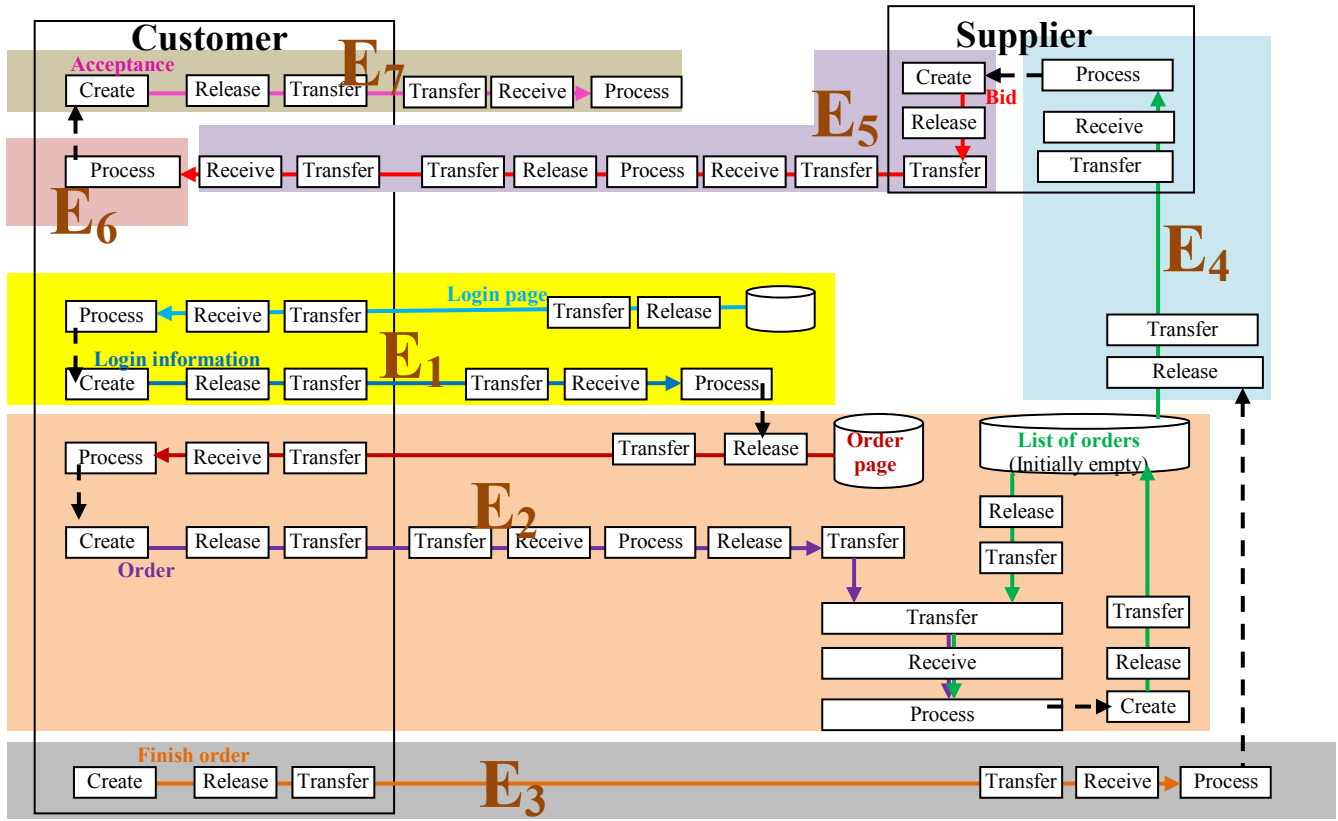

**Figure 27. The events in the TM model of the Submit Order scenario.**

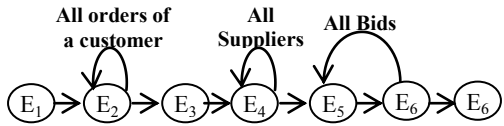

**Figure 28. The behavior in the TM model of the Submit Order scenario.**

The paper's many examples show TM's viability as a conceptual modeling tool and as an instrument to analyze functionality. We view this study as a starting point that could lead to a different approach for analyzing functionality. Exploring the TM's potential in this direction can point in many directions. To demonstrate, we will focus on two specific issues of how TM clarifies problems related to functions.

### A. Basic Functions

We already traced the notion of function and its elementary types based on the five TM operations: create, process, release, transfer, and receive. Thus, as a description of a function, "to do" is divided into types of doing: to create, to process, to release, to transfer, and to receive. Such a contribution allows for deeper analysis of the notion of function.

Contrasting the TM approach with current attempts to find basic functions, as in [9], defines a function as basic if its final state "is a single fact".

Basic functions are elementary functions that, in functional hierarchy, always appear in the lowest position and as non-decomposable. According to [9], the goal of the basic function "to deliver goods to Rome" comprises "the goods are located in Rome." Reference [9]'s example can be represented in semi-formal language as:

Goods: Rome.transfer.receive.store

Thus, TM modeling exposes more basic functions. In [9]'s example, "to deliver" involves the elementary operations "to arrive" (transfer) and "to receive". A thing could arrive but never be received. For example, an e-mail could arrive at the communication port but, because of some error, never reach the input buffer. Goods can arrive in Rome but never be received because they are stolen upon arrival. The interesting problem on the relationship between "function" and "goal" remains to be explored in future research.

### B. Function Requirement

According to [9], function requirements are intentional entities referring to a "chunk of reality" that should be present to realize the function. For example, realizing the function "to hammer nails" requires nails and a physical object to which the nails should be hammered.

Fig. 29 shows the TM representation of this example, which clarifies the "chunk of reality which should be present if the function is to be realized." This includes a hand that processes a hammer to make a nail penetrate the physical object. Note that the existence of the hand, hammer, and nails (i.e., their creation) is not shown for simplicity sake. To give the dynamic picture of hammering nails, Figs. 30 and 31 provide its events and behavior.

We can verify that extensive applicability to many such problems related to functionality can be explored further using TMs.

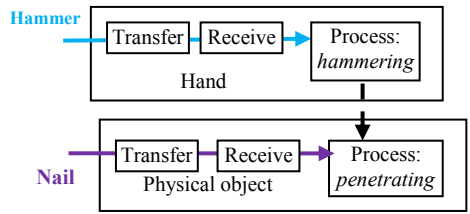

**Figure 29. TM representation of "to hammer nails".**

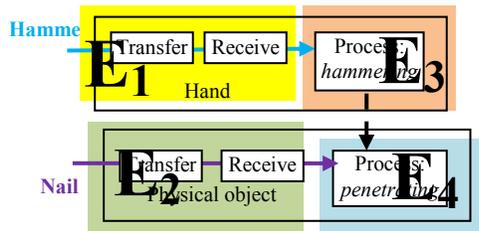

**Figure 30. Events of "to hammer nails".**

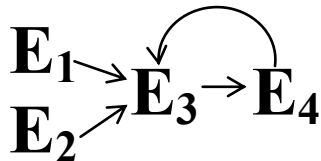

**Figure 31. The behavior of "to hammer nails".**